\documentstyle[12pt]{article}
\font \bigrm = cmr17 at 24pt
\font \secfont = cmbx12 at 16pt
\def \lesssim
{\,\lower 3 pt \hbox{\vbox{\hbox{$<$} \kern -9 pt \hbox{$\sim$}}}\,}
\def \greatersim
{\,\lower 3 pt \hbox{\vbox{\hbox{$>$} \kern -9 pt \hbox{$\sim$}}}\,}
\voffset = - 0.75 in
\def \MSbar {\vbox{\hrule\kern 1pt\hbox{\rm MS}}}
\def \GeV { {\ \rm GeV} }

\def \DESsection #1 {\bigskip\leftline{\secfont #1}\par\medskip
\noindent}
\catcode`\@=11
\def\@citex[#1]#2{\if@filesw\immediate\write\@auxout
	{\string\citation{#2}}\fi		
\def\@citea{}\@cite{\@for\@citeb:=#2\do		
	{\@citea\def\@citea{,}\@ifundefined		
	{b@\@citeb}{{\bf ?}\@warning
	{Citation `\@citeb' on page \thepage \space undefined}}
	{\csname b@\@citeb\endcsname}}}{#1}}

\newif\if@cghi
\def\cite{\@cghitrue\@ifnextchar [{\@tempswatrue
	\@citex}{\@tempswafalse\@citex[]}}
\def\citelow{\@cghifalse\@ifnextchar [{\@tempswatrue
	\@citex}{\@tempswafalse\@citex[]}}
\def\@cite#1#2{{$\null^{#1)}$\if@tempswa\typeout
	{IJCGA warning: optional citation argument 
	ignored: `#2'} \fi}}

\def\red#1{#1}
\def\blue#1{#1}
\def\green#1{#1}

\input epsf.tex
\def\DESepsf(#1 width #2){\epsfxsize=#2 \epsfbox{#1}}
\begin{document}
%
\null
\vskip 4 true cm
\centerline{\bigrm JET OBSERVABLES}
\medskip
\centerline{\bigrm IN THEORY AND REALITY}
\bigskip
\centerline{Davision E.\ Soper}
\centerline{\it Institute of Theoretical Science}
\centerline{\it University of Oregon}
\centerline{\it Eugene, Oregon 97403 USA}
\vskip 3 cm
\centerline{Abstract}
I discuss the one jet inclusive jet cross section, $d\sigma/dE_T$
emphasizing the concept of infrared safety and the cone definition of
jets. Then I estimate the size of power corrections to the jet cross
section, which become important at smaller values of
$E_T$.

\vskip 1 true cm
\leftline{Talk at the Rencontre de Moriond, QCD Session}
\leftline{Les Arcs, France,\ March 1997}
\pagebreak
%
\DESsection{Jet definitions and infrared safety}

Jets are an obvious feature of the final state for
$p + \bar p \to hadrons$. Although the jets are obvious, for a
quantitative comparison of theory and experiment, one must carefully
specify what one means by a jet. The jet definition associates with
each jet a transverse energy $E_T$,  a rapidity
$\eta$,  and an azimuthal angle $\phi$. Then one can form the
inclusive jet cross section
\begin{equation}
{ d \sigma \over d E_T}
= { 1 \over \eta_2-\eta_1}\int_{\eta_1}^{\eta_2}\!\! d \eta 
\int_0^{2 \pi}\!\! d \phi\ 
{ d \sigma \over d E_T d \eta d \phi}.
\label{xsect}
\end{equation}
With a suitable definition the cross section has the form
\begin{equation}
{\green{d \sigma \over d E_T\, d\eta\, d \phi}} \approx\
\sum_{a,b} \int_{x_A}^1\! d \xi_A \int_{x_B}^1\! d \xi_B\
{\blue{f_{a/A}(\xi_A,\mu)}}\ {\blue{f_{b/B}(\xi_B,\mu)}}\
{\red{{{d \hat\sigma^{ab}(\mu) \over d E_T\, d\eta\, d \phi}}}},
\end{equation}
where the $f_{a/A}(\xi,\mu)$ are the parton distributions and where
${\red{{{d \hat\sigma^{ab} / [d E_T\, d\eta\, d \phi]}}}}$
is the perturbatively calculated hard scattering function. (The
lowest order is $\alpha_s^2$; current calculations\cite{EKS,Jetrad}
reach $\alpha_s^3$.)

The cross section should be insensitive to long times $\Delta t \gg
1/E_T$. However, we know that, long after the hard interaction, soft
particles are emitted or absorbed. Thus, $d \sigma / dE_T$ should be
insensitive to soft particles. We know also that long after the hard
interaction a fast particle can split into two collinear particles or
two collinear fast particles can join. Thus $d \sigma / dE_T$ should
be insensitive to collinear splitting and joining. A measureable
quantity such as a jet cross section that has these properties is
said to be infrared safe.

I consider here the ``Snowmass Accord'' jet definition\cite{snowmass}.
Define a jet cone of radius R in $\eta$-$\phi$ space. Typically, $R =
0.7$. The jet variables $E_{T,J},\eta_J,\phi_J$ are defined in terms
of the particles by
\begin{equation}
E_{T,J} = \sum_{i\in {\rm cone}}E_{T,i}\,,
\end{equation}
\begin{equation}
\phi_J = {1 \over E_{T,J}}\sum_{i\in {\rm cone}}E_{T,i}\ \phi_i\,,
\hskip 1 cm
\eta_J = {1 \over E_{T,J}}\sum_{i\in {\rm cone}}E_{T,i}\ \eta_i\,.
\end{equation}
\medskip
The cone axis must agree with the jet axis $(\eta_J,\phi_J)$.
One iterates until agreement is reached.

There are some difficulties that require one to amend this jet
definition.\cite{CDFdef,D0def} First, sometimes the cones overlap. If
there is more than some specified amount of transverse energy  in the
overlap region, then one merges the jets. Otherwise, one splits the
transverse energy in the overlap region between the two jets.
Second, there will be transverse energy in the jet cone that is not
related to the high $P_T$ parton that made the jet but instead is
associated with the underlying event.  Accordingly, one subtracts
$\rho\,\pi R^2$ from $E_T$, where $\rho$ is the average $E_T$ per unit
$d \eta\, d \phi$ in minimum bias events. At the Fermilab Tevatron,
$\rho\,\pi (0.7)^2 \approx 1.1 \GeV$. Note that one does {\it not} do
this in the corresponding order $\alpha_s^3$ theoretical calculation
since the underlying event is not part of the calculation.

In the theoretical calculation, one often modifies\cite{lookinside}
the Snowmass defintion to restrict merging partons into jets
according to a variable $R_{\rm sep}$. If two partons have
\begin{equation}
(\eta_1 - \eta_2)^2 +  (\phi_1 - \phi_2)^2 
> R_{\rm sep}^2 \,,
\end{equation}
one does {\it not} merge them, even though
\begin{equation}
(\eta_i - \eta_{\rm jet})^2 +  (\phi_i - \phi_{\rm jet})^2 
< R^2 ,
\hskip 1 cm i = 1,2.
\end{equation}
Why? Studies\cite{lookinside} of the $E_T$ distribution within jets
suggest that the experimental jet algorithms fail to merge subjets
that are rather far separated. The value $R_{\rm sep} = 1.3 R$ is
suggested. This lowers the theoretical cross section about 4\%.

The theory for jet production comes with theoretical uncertainties.
There are uncalcuated $\alpha_s^4$ and higher perturbative
contributions. A typical estimate\cite{EKS} of this uncertainly is 
$\pm 15\%$. In addition, there is uncertainty in the parton
distributions. One may guess $\pm 20\%$ for this uncertainty for
$E_T < 300 \GeV$. The uncertainty is presumably larger for $E_T >
300 \GeV$\  because the gluon distribution is largely
unknown\cite{CTEQglue} at large $x$. Finally, there are corrections
suppressed by a power of ``1 GeV''/$E_T$. In the remainder of this
talk, I attempt to estimate these corrections using simple models.

\DESsection{Power suppressed corrections}

In this section, I address power suppressed corrections to the
theory. I will present graphs of {\it (Data - Theory ) /Theory}.
Both data and theory refer to the jet cross section $d\sigma/dE_T$
for $0.1 < |\eta| < 0.7$ and $R = 0.7$ as in Eq.~(\ref{xsect}). The
data is from CDF\cite{CDF1800} and has been corrected by CDF by
subtracting from the jet $E_T$ an estimate of the transverse energy
contained in the jet cone that arises from the underlying event
($\Delta E_T
\approx$ 1.1 GeV). The theory here is straightforward order
$\alpha_s^3$ QCD with CTEQ4M partons. The theoretical calculation
uses an $R_{\rm sep}$ correction to the Snowmass algorithm, with
$R_{\rm sep} = 1.3 R$.

{\it Splash in}.
In $\alpha_s^3$ theory, the transverse energy of a jet comes from one
or two partons. It is the decay products of these partons (using the
picture embedded in Monte Carlo models), that one wants to capture
in the jet cone. However, other particles can splash into the cone.
The underlying event creates soft particles that will get into the
cone. An estimate of this effect has already been subtracted from
$E_T$ in the data, so we do not consider it further. In addition, the
accelerated initial state partons radiate soft gluons, whose decay
products can splash into the cone. Let us try to estimate this effect.

If transverse energy $\Delta  E_T^{\rm in}$ is added to the jet
$E_T$, the cross section should change by 
\begin{equation}
{ d \sigma \over d E_T} \approx
\left( d \sigma \over d E_T \right)_{\!\rm NLO}
\left\{ 1 + n\
{ \Delta E_T^{\rm in} \over E_T}
\right\},
\label{ETin}
\end{equation}
where
\begin{equation}
n(\ln(E_T)) = - { d \over d\ln(E_T)}\,
\ln\left[\left( d \sigma \over d E_T \right)_{\!\rm NLO}
\right].
\end{equation}
The factor $1/E_T$ makes this a small correcton for large $E_T$.
However, $n$ is large.

Marchesini and Webber\cite{Marchesini} have suggested a method for
estimating from data the level of transverse energy in jet events
without including the $E_T$ from the third of three partons in a
perturbative calculation. Lacking such a data-based analysis, I use
Marchesini and Webber's Monte Carlo study for the amont of extra
splash-in: 
\begin{equation}
\Delta E_T^{\rm in} \approx  0.6 \GeV.
\end{equation}

{\it Splash-out}. Some of the partonic transverse energy can leak out
of the jet cone. The order $\alpha_s^3$ perturbation theory gets this
effect partly right: in a three parton final state the third
parton can escape the jet cone. However, using the picture embedded
in Monte Carlo models, the late stages of partonic branching and the
final hadronization of the partons can also result in transverse
energy escaping the jet cone.  Here is a simple model for this effect.

Consider the hadrons that represent the decay products of a high
$E_T$ parton. Let $\eta$ be the rapidity of the hadrons relative
to jet axis. Let $\vec k_T$ be the transverse momentum of the
particles relative to jet axis. Let the distribution of hadrons be
\begin{equation}
{ d N \over d \eta d\vec k_T} =
{ A \over \pi \langle k_T^2\rangle}\
\exp\left\{ 
- k_T^2/\langle k_T^2\rangle
\right\},
\end{equation}
where $A$ is the number of hadrons per unit rapidity and
$\langle k_T^2\rangle$ is average $k_T^2$ of the hadrons.
Then the $E_T$ lost is approximately
\begin{equation}
E_T^{\rm out} = 
\int_0^{\eta_1}\!\! d \eta \int\! d \vec k_T\
{1 \over 2} |\vec k_T | e^\eta \
{ d N \over d \eta d\vec k_T},
\end{equation}
where 
$
\eta_1 = - \ln\left(\tan(R/2)\right).
$
Performing the integral gives
\begin{equation}
E_T^{\rm out} = 
{ \sqrt \pi \over 4} A \sqrt{\langle k_T^2\rangle}
\left(e^{\eta_1} - 1
\right).
\end{equation}
Taking $\sqrt{\langle k_T^2\rangle}$ = 0.3 GeV and\cite{Zhou} $A$ =
5, I find
\begin{equation}
E_T^{\rm out} \approx 1.1 \GeV.
\end{equation}
This estimate can be used to correct the theoretical cross section,
using the analog of Eq.~(\ref{ETin}) (with the opposite sign for 
$E_T^{\rm out}$ instead of $E_T^{\rm in}$).

In Fig.~\ref{corrections} below, I show the data compared to theory.
In the same graph, I show a curve representing ({\it Modified Theory
$-$ Theory) / Theory}, where the modified theory includes a
correction for splash-in/splash-out. Evidently, the correction does
not improve the agreement between theory and experiment. The size of
the correction provides some estimate of the theory error from this
source.
\begin{figure}[htb]
\centerline{\DESepsf(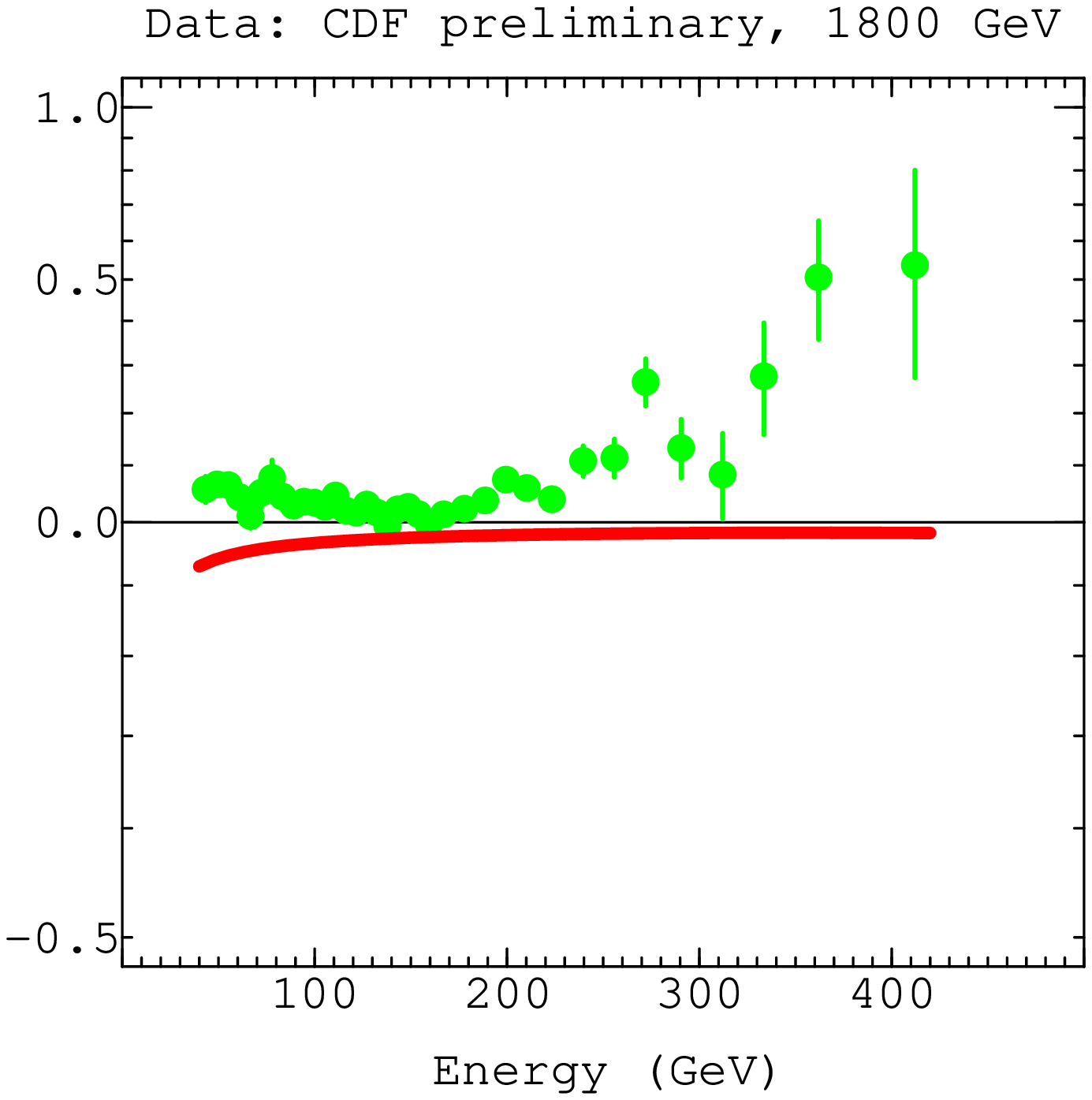 width 8 cm)
\DESepsf(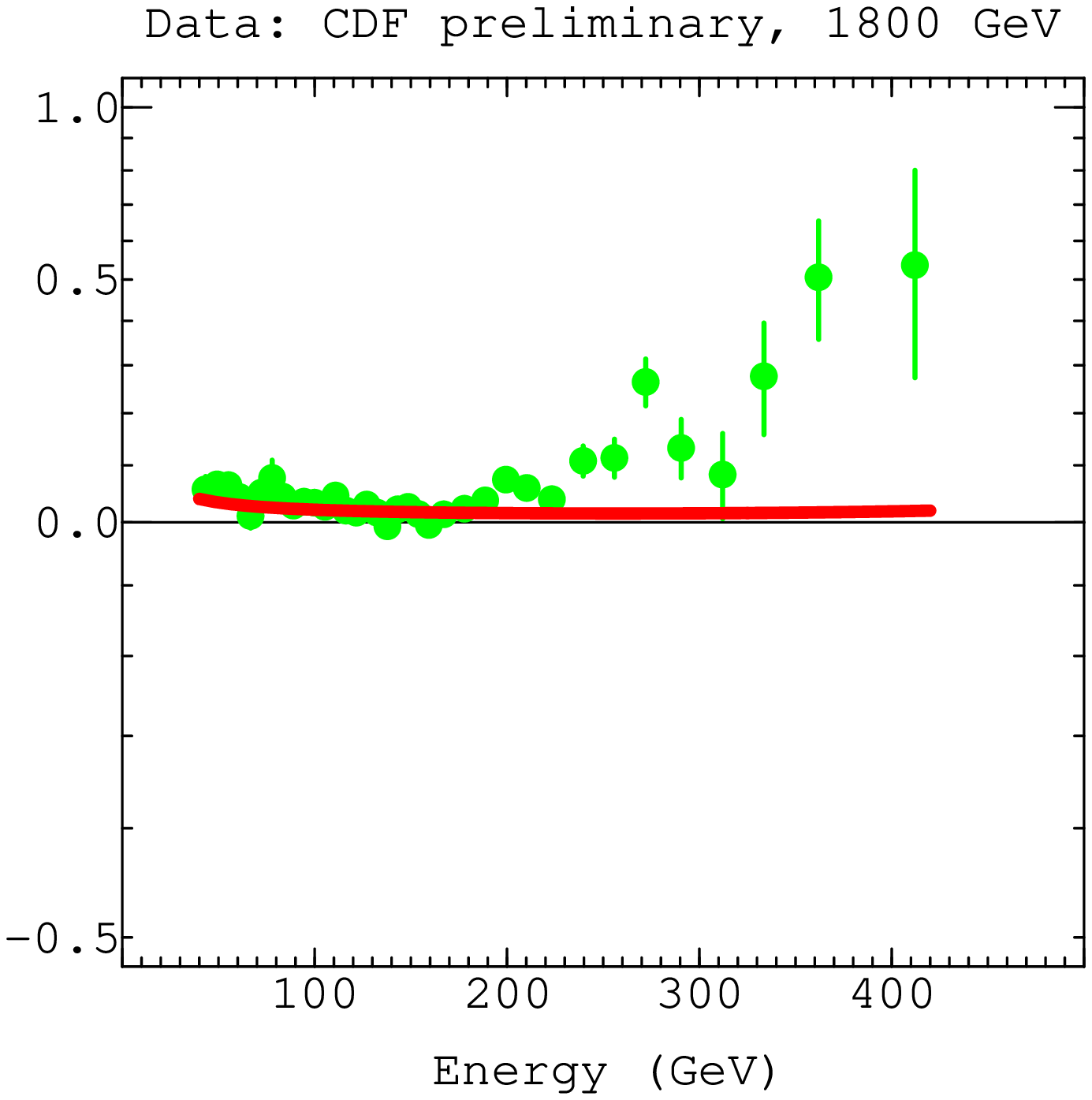 width 8 cm)}
\caption{$(Data - Theory)/Theory$ for the jet cross section
versus transverse energy $E_T$. In the left hand graph, the curve
shows the correction to the theory for splash-in/splash-out.
In the right hand graph, the curve
shows the correction to the theory for $k_T$ smearing.}
\label{corrections}
\end{figure}

{\it Transverse momentum smearing.} In an order $\alpha_s^3$
calculation, incoming partons have zero transverse momentum.
Thus the observed jet can recoil against only one or two jets.
But in a more realistic model, the incoming partons can radiate
multiple soft gluons, each carrying some transverse momentum. In
addition, the partons in the proton wave function can have a 
``primordial $k_T$'' by virtue of being bound in a hadron of finite
size.

Let us try a simple model that accounts for these effects. There is a
risk of double counting, since we may count the same gluon both as
one of the multiple soft gluons and as one of the final state gluons
in the hard scattering. We simply hope that the double counting
effect will be small.

Define a function $G(\vec P_T)$ that represents the measured cross
section for one jet inclusive production:
\begin{equation}
G(\vec P_T) = { d \sigma \over d \vec P_T}
= { 1 \over 2 \pi P_T}\ { d \sigma \over d P_T},
\hskip 1.0 cm P_T \equiv E_T.
\end{equation}
Let $G\!\left(\vec P_T\right)_{\rm NLO}$ be the same cross section
calculated at next to leading order. Then we can make a simple model
for $G(\vec P_T)$:
\begin{equation}
G(\vec P_T) = \int\! d\vec k_T\ \rho(\vec k_T; P_T) \
G\!\left(\vec P_T - {1 \over 2}\,\vec k_T\right)_{\rm NLO}
,%
\end{equation}
where $\rho(\vec k_T; P_T)$ is a smearing function that represents
the probability that the partons entering the hard scattering had
transverse momentum $\vec k_T$. This function depends on $P_T$
because large $P_T$ in the hard scattering means more soft gluon
radiation.

Supposing that $k_T \ll P_T$, we have
\begin{eqnarray}
G(\vec P_T)&\approx&\int\! d\vec k_T\ \rho(\vec k_T; P_T) \
\biggl\{
G(\vec P_T)_{\rm NLO}
\nonumber\\
&&- {1 \over 2} k_T^i \nabla_i G(\vec P_T)_{\rm NLO}
+ { 1 \over 8}k_T^i k_T^j \nabla_i \nabla_j G(\vec P_T)_{\rm NLO}
\biggr\}.
\end{eqnarray}
Define
\begin{equation}
n(\ln(P_T)) = - { d \over d\ln(P_T)}\,
\ln\left(2\pi P_T\ G(\vec P_T)_{\rm NLO}
\right).
\end{equation}
Then
\begin{equation}
{ d \sigma \over d E_T} \approx
\left( d \sigma \over d E_T \right)_{\!\rm NLO}
\left\{ 1 + { (n+1)^2 - n' \over 16}\
{ \langle k_T^2\rangle \over E_T^2}
\right\}.
\end{equation}

To obtain a quantitative estimate, I use $n(\ln(E_T))$ from a NLO
calculation and
\begin{equation}
\langle k_T^2\rangle
\approx
2 m E_T,
\hskip 1 cm m = 0.35 \GeV.
\end{equation}
This formula is a rule of thumb. The number for $m$ is based on
identifying $2E_T$ with $M_Z$ in $p + \bar p \to Z + X$, supposing
that the $k_T$ distribution of the $Z$ is approximately gaussian for
$k_T \ll M_Z$, and identifying $\langle k_T^2\rangle$ by noting
that\cite{zkt} $d \sigma /d k_T$ peaks at $k_T \approx 4 \GeV$. Then
\begin{equation}
{ d \sigma \over d E_T} =
\left( d \sigma \over d E_T \right)_{\!\rm NLO}
\left\{ 1 + { (n+1)^2 - n' \over 8}\
{ m \over E_T}
\right\}.
\label{kt}
\end{equation}
(There is computer code by Baer and Reno\cite{BaerReno} that does
this sort of smearing for direct photon production, using a Monte
Carlo style calculation.) In Fig.~\ref{corrections}, I show the data
compared to NLO theory with a correction for $k_T$ smearing shown as
a curve.

\begin{figure}[htb]
\centerline{\DESepsf(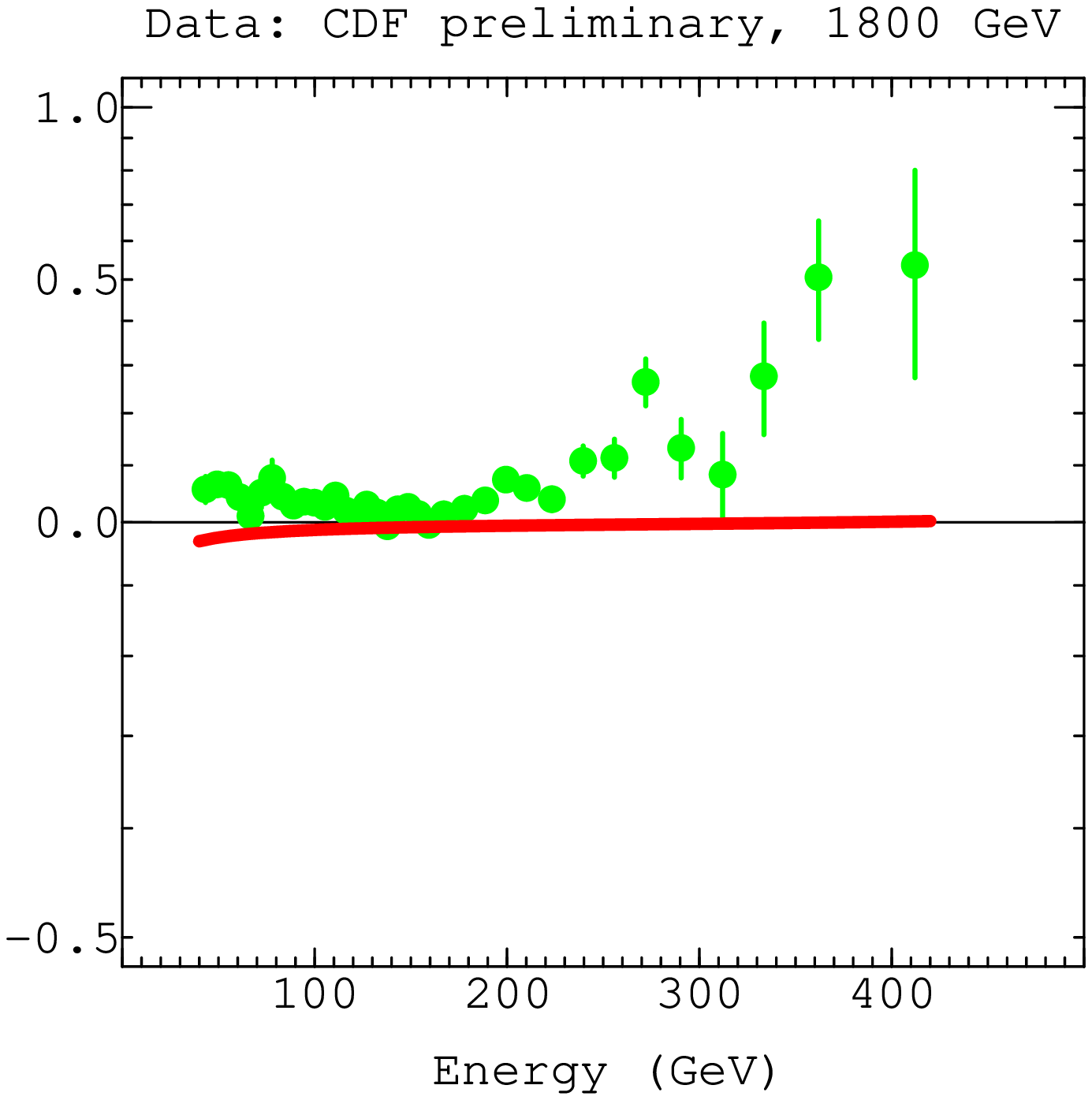 width 8 cm)
\DESepsf(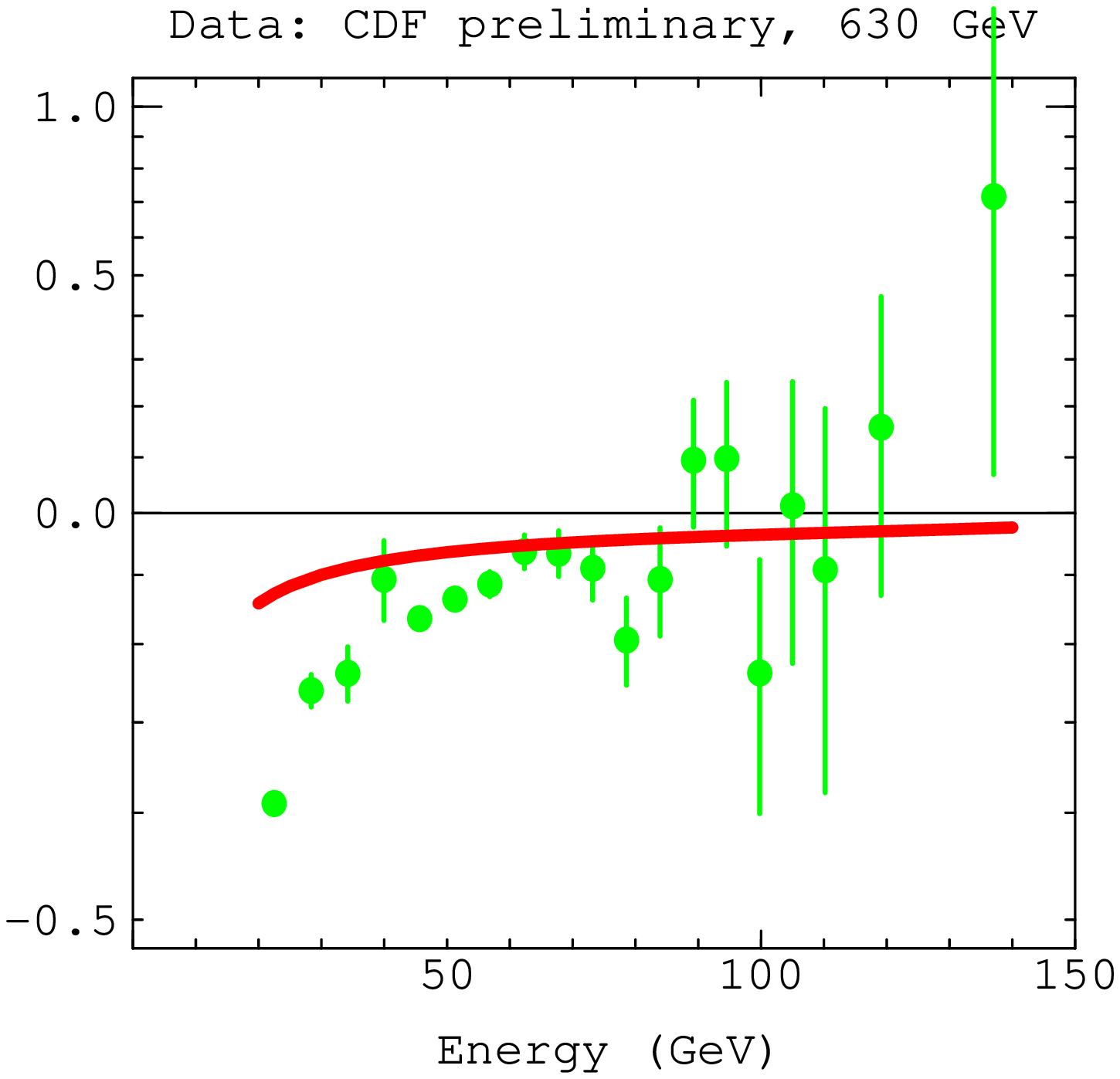 width 8 cm)}
\caption{Graph of $(Data - Theory)/Theory$ versus $E_T$ for
with curves showing the net power
supressed correction as estimated using Eqs.~(\protect\ref{ETin}) and
(\protect\ref{kt}). The left hand graph is for $\protect\sqrt s =
1800 \GeV$ while the right hand graph is for $\protect\sqrt s = 630
\GeV$}
\label{netgraphs}
\end{figure}

{\it The net result.} In Fig.~\ref{netgraphs}, I show the data
compared to the  NLO theory with the net correction for
splash-in/splash-out and $k_T$ smearing shown as a curve.

The net correction is quite small. If we take the size of the various
corrections as an error estimate, we can estimate a theory error from
power suppressed corrections of about $\pm 10\%$ at $E_T = 40 \GeV$
and $\pm 4\%$ at $E_T = 200 \GeV$.

In Fig.~\ref{netgraphs}, I also show the CDF data\cite{CDF630} at
$\sqrt s = 630 \GeV$ compared to the  NLO theory with the net
correction for splash-in/splash-out and $k_T$ smearing shown as a
curve. I use the same parameters as for $\sqrt s = 1800 \GeV$ except
that (estimating from the Monte Carlo study used
earlier\cite{Marchesini}), I reduce the splash-in parameter
$E_T^{\rm in}$ from 0.6 GeV to 0.3 GeV.

If we take the size of the various corrections as an error
estimate, then the estimated error is larger than at $\sqrt s = 1800
\GeV$, perhaps $\pm 20\%$ at $E_T = 20 \GeV$. It may be significant
that, while the net effect at $\sqrt s = 1800 \GeV$ is really quite
small, the net effect at $\sqrt s = 630 \GeV$ is not so small, and
is in the right direction to improve the agreement between theory
and experiment.

\bigskip
I thank Anwar Bhatti of the CDF Collaboration for help with the
CDF data.

\end{document}